\documentclass[aps,prd,nofootinbib,superscriptaddress,floatfix,nofootinbib,amsmath,amssymb]
{revtex4}
\usepackage[dvips]{graphicx}

\begin{document}

\begin{flushright}
USTC-ICTS-10-02
\end{flushright}

\title{Cosmic age test in inhomogeneous cosmological models mimicking $\Lambda$CDM on the light cone}

\author{Mi-Xiang Lan}
\email{lan@mail.ustc.edu.cn} \affiliation{Department of Modern
Physics, University of Science and Technology of China, Hefei
230026, China}  \affiliation{Key Laboratory of Frontiers in
Theoretical Physics, Institute of Theoretical Physics, Chinese
Academy of Sciences, Beijing 100190, China}

\author{Miao Li}
\email{mli@itp.ac.cn} \affiliation{Kavli Institute for Theoretical
Physics China, Chinese Academy of Sciences, Beijing 100190, China}
\affiliation{Key Laboratory of Frontiers in Theoretical Physics,
Institute of Theoretical Physics, Chinese Academy of Sciences,
Beijing 100190, China}

\author{Xiao-Dong Li}
\email{renzhe@mail.ustc.edu.cn} \affiliation{Department of Modern
Physics, University of Science and Technology of China, Hefei
230026, China} \affiliation{Interdisciplinary Center for Theoretical
Study, University of Science and Technology of China, Hefei 230026,
China} \affiliation{Key Laboratory of Frontiers in Theoretical
Physics, Institute of Theoretical Physics, Chinese Academy of
Sciences, Beijing 100190, China}

\author{Shuang Wang}
\email{swang@mail.ustc.edu.cn} \affiliation{Department of Modern
Physics, University of Science and Technology of China, Hefei
230026, China} \affiliation{Key Laboratory of Frontiers in
Theoretical Physics, Institute of Theoretical Physics, Chinese
Academy of Sciences, Beijing 100190, China}

\begin{abstract}
The possibility of reconstructing a spherically symmetric
inhomogeneous Lema\^{\i}tre-Tolman-Bondi (LTB) model with
$\Lambda$CDM observations has drawn much attention. Recently, an
inhomogeneous LTB model having the same luminosity-distance and
light-cone mass density of the homogeneous $\Lambda$CDM model was
reconstructed. From the Wilkinson microwave anisotropy probe 7-year
measurements together with other cosmological observations, we
calculate the cosmic age at our position in this LTB model, and
obtain a constraint $t_{LTB}<11.7$Gyr at 1$\sigma$ confidence level.
We find that this result is, although 2Gyr younger compared with the
age of the homogeneous $\Lambda$CDM model, still within 1$\sigma$
agreement with the constraint of cosmic age given by current
astronomical measurements. We expect that in the future with the
help of more advanced observations we can distinguish the
reconstructed inhomogeneous LTB model from the homogeneous
$\Lambda$CDM model.
\end{abstract}

\maketitle

\section{Introduction}\label{sec:intro}

The problem of dark energy has become one of the most important
issues of the modern cosmology since the observations of type Ia
supernovae (SNe Ia) \cite{Riess:1998cb} first indicated that the
universe is undergoing an accelerated expansion at the present stage
(if assuming that the universe is described by the
Friedmann-Lema\^{i}tre-Robertson-Walker (FLRW) model). Many
cosmologists believe that the identity of dark energy is the
cosmological constant which fits the observational data very well.
However, one still has reasons to dislike the cosmological constant
since it suffers from the theoretical problems such as the
``fine-tuning'' and the ``cosmic coincidence'' puzzles \cite{DErev}.
Thus, a variety of proposals for dynamic dark energy have emerged.
For example, the ``scalar field" model \cite{quintenssence} has been
explored for a long time. Besides, the ``holographic dark energy"
models \cite{HDE}, which arise from the holographic principle of
quantum gravity theory, has also attracted much attention.

There are also some other theoretical approaches to explain the
current cosmic acceleration. For example, it is argued that the
acceleration of the universe may signify the breakdown of Einstein's
theory of general relativity \cite{fR}. Another interesting idea
\cite{Void} is based on the assumption that the universe is
described by the spherically symmetric, inhomogeneous
Lema\^{\i}tre-Tolman-Bondi (LTB) metric \cite{LTB1} \cite{LTB2}
\cite{LTB3}. Recently, some authors further proposed a possibility
of mimicking the cosmological constant in an inhomogeneous universe
\cite{Mustapha} \cite{CBKH} \cite{Kolb} \cite{AER}. The idea is
that, since cosmological observations are limited on the light cone,
it is possible to reconstruct an inhomogeneous cosmological model
(indistinguishable from the homogeneous $\Lambda$CDM model) to
explain the cosmic acceleration without a cosmological constant. The
formalism of reconstruction was developed in \cite{Mustapha} and was
applied to the $\Lambda$CDM model in \cite{CBKH}. In \cite{Kolb},
the authors constructed a spherically symmetric, inhomogeneous
cosmological model reproducing the luminosity-distance and the
light-cone mass density of $\Lambda$CDM model up to $z=2$.

In this work, we focus on the reconstructed inhomogeneous
cosmological model and investigate whether it is consistent with
cosmological observations. Since in this model $\Lambda$CDM
observations are exactly reconstructed on the light-cone, we should
seek for some observational test \textit{not limited on the
light-cone} to distinguish it from the standard $\Lambda$CDM model.

Fortunately, we find that the age of the universe is an appropriate
touchstone. The cosmic age, which depends on the evolution of the
universe at a comoving position, contains information not limited on
the light-cone (the age is uniform in the $\Lambda$CDM model but may
be dependent on the position in an inhomogeneous cosmological
model). Thus it may reveal the discrepancies between the
$\Lambda$CDM model and the reconstructed inhomogeneous model. On the
other hand, it is rather convenient to use the cosmic age to test
the validity of a specific cosmological model. To do this one may
just compare the result with the age of some old objects in our
universe \cite{Age}. For example, to be consistent the age of our
universe in our position must not be younger than the age of the
oldest stellar in the Milky Way.

This paper is organized as follows. In Sec. \ref{sec:LTB}, following
the procedure of \cite{Kolb}, we introduce the inhomogeneous LTB
model and explain how to reconstruct $\Lambda$CDM observations in
this model. In Sec. \ref{sec:CosmicAge}, we calculate the cosmic age
at our position and compare the result with the age of some old
objects. We summarize in Sec. \ref{sec:Summary}.

\section{LEMA\^{I}TRE-TOLMAN-BONDI MODELS}\label{sec:LTB}

In this section, following the procedure of \cite{Kolb}, we explain
how to reconstruct an inhomogeneous LTB model having the same
luminosity distance and light-cone mass density of the homogeneous
$\Lambda$CDM model.

The LTB models are spherically symmetric cosmological solutions to
the Einstein equations with a dust stress-energy tensor. The general
metric for the LTB models in a synchronous comoving coordinate takes
the form,
\begin{equation}
ds^2=-dt^2+\frac {R^{\prime 2}(r,t)} {1+\beta(r)} dr^2 +
R^2(r,t)d\Omega^2.
\end{equation}
Following \cite{Mustapha,Kolb}, we use the prime superscript to
denote $\partial/\partial r$, and the overdot to denote
$\partial/\partial t$. Noticing that here $r$ is a dimensionless
coordinate, while $R(r,t)$ has the dimension of length. The
Robertson-Walker metric can be recovered after performing
$R(r,t)\rightarrow a(t)r$ and $\beta(r)\rightarrow -kr^2$. Solving
the Einstein Equations one obtains the generalized ``Friedmann
Equations" for $R(r,t)$ and $\rho(r,t)$,
\begin{equation}
\label{E1}
\dot{R}(r,t)=\sqrt{  \beta(r)   +   \frac{\alpha(r)}{R(r,t)}  },
\end{equation}
\begin{equation}
\label{E2} \kappa \rho(r,t)  =  \frac  {\alpha^{\prime}(r)}
{R^{2}(r,t)R^{\prime}(r,t)}.
\end{equation}
And the photon radial null geodesic equation for $\hat{t}(r)$ is
found directly from the LTB metric (we are only interested in the
past light cone),
\begin{equation}
\frac {d\hat{t}(r)} {dr} = - \frac {R^{\prime}(r,\hat{t}(r))}
{\sqrt{1+\beta(r)}}.
\end{equation}
For convenence we denote quantities on the light cone by a hat. Thus
we have,
\begin{equation}
R(r,\hat{t}(r))\equiv\hat{R};  \ \ \
R^{\prime}(r,\hat{t}(r))\equiv\hat{R^{\prime}}; \ \ \
\rho(r,\hat{t}(r))\equiv\hat{\rho}.
\end{equation}
Then let us focus on the reconstruction procedures. The method was
discussed by Mustapha, Hellaby, and Ellis in 1997 \cite{Mustapha},
and was recently applied to the $\Lambda$CDM model in \cite{CBKH}
\cite{Kolb} (A related formalism was developed by \cite{CR}
\cite{YKN}). Following their procedure, we take advantage of a
coordinate freedom and rescale $r$ so that \emph{on the light cone},
\begin{equation}
\hat{R}^{\prime}=H^{-1}_{0}\sqrt{1+\beta(r)}.
\end{equation}
The corresponding coordinate transformation is
$dr_1=\frac{H_0\partial \hat R/\partial r }{\sqrt{1+\beta(r)}}dr$.
In the following, for simplicity we will use $r$ to denote $r_1$.
The redshift takes the form (see Sec. 2.3 in \cite{Mustapha})
\begin{equation}
\frac{d\hat z(r)}{dr}=(1+z)\frac{ \hat {\dot
{R^\prime}}}{\sqrt{1+\beta(r)}}.
\end{equation}
Following the procedure of \cite{Kolb}, we reconstruct
$\hat{d}_L(z)$ and $\hat{\rho}(z)$ of the $\Lambda$CDM model. To do
this we require that on the light cone,
\begin{equation} \label{reconstruction}
(1+z)^2\hat R(z)=(1+z)\int^z_0 \frac {dz_1}
{H_{\Lambda\textrm{CDM}}(z_1)}, \ \ \
\hat{\rho}(z)dV_{LTB}=\rho_{M,\Lambda\textrm{CDM}}dV_{\Lambda\textrm{CDM}},
\end{equation}
where $\rho_{M,\Lambda\textrm{CDM}}(z)$ stands for mass density in
the $\Lambda$CDM model, and $H_{\Lambda\textrm{CDM}}(z)$ stands for
the Hubble constant in the $\Lambda$CDM model. These quantities take
the forms,
\begin{equation} \label{LCDMValue}
\rho_{\Lambda\textrm{CDM}}(z)=3\Omega_mM_p^2H_{\Lambda\textrm{CDM}}(z)^2,\
\
H_{\Lambda\textrm{CDM}}=H_0\sqrt{\Omega_m(1+z)^3+\Omega_{\Lambda}}.
\end{equation}
For simplicity we only consider a flat $\Lambda$CDM model. We use
the notations,
\begin{equation}
\Omega_{m}=\frac{\rho_{m}(0)}{\rho_{C}(0)},
\ \ \ \Omega_{\Lambda}=\frac{\rho_{\Lambda}(0)}{\rho_{C}(0)},\ \ \ \rho_C=3M_p^2H_0^2,\ \ \ \Omega_m+\Omega_{\Lambda}=1 .
\end{equation}
Then from Eq. (\ref{reconstruction}) it is straightforward to derive
the following expressions,
\begin{equation} \label{BasicEq.1}
\hat{R}(z)= \frac {1}{1+z} \int^z_0 \frac
{dz_1}{H_{\Lambda\textrm{CDM}}(z_1)},
\end{equation}
\begin{equation} \label{BasicEq.2}
H^{-1}_0\hat{R}^2(z)\kappa\hat{\rho}(z) \frac{dr}{dz} =
\frac{3\Omega_{m}H^2_0}{H_{\Lambda
\textrm{CDM}}(z)}\big[\int^z_0\frac{dz_1}{H_{\Lambda
\textrm{CDM}}(z_1)}\big]^2 .
\end{equation}
Furthermore, the following three equations can be obtained by
solving the corresponding $z$, $\alpha(r)$ and $\beta(r)$ (Eqs.
(19-21) in \cite{Kolb})
\begin{equation}\label{BasicEq.3}
\frac {dz}{dr} =(1+z) \frac {H_{\Lambda \textrm{CDM}}(z)}{H_0} ,
\end{equation}
\begin{equation}\label{BasicEq.4}
\frac {d\alpha}{dr}=\frac
{1}{2}H^{-1}_0\hat{R}^2(z)\kappa\hat{\rho}(z)\big[\frac{1}{H_0d\hat{R}/dr}\big(1-\frac
{\alpha}{\hat{R}}\big) + H_0 \frac {d\hat{R}}{dr}\big],
\end{equation}
\begin{equation}\label{BasicEq.5}
\beta(r)=(\frac{d\alpha}{dr}\frac{1}{H^{-1}_0\hat{R}^2\kappa\hat{\rho}})^2-1.
\end{equation}
It should be emphasized that for these values of $\alpha(z)$ and
$\beta(z)$, the corresponding LTB model \emph{exactly reproduces}
the luminosity-distance relation and light-cone mass density of the
$\Lambda$CDM model. For convenience, in the following context we
will use ``LTB-$\Lambda$CDM model" to denote this reconstructed LTB
model.

It should be stressed that since the LTB-$\Lambda$CDM model is
reconstructed by mimicking $\Lambda$CDM model, it has the same
luminosity-distance-redshift relation and light-cone
mass-density-redshift relation as the homogeneous $\Lambda$CDM
model. Thus when estimating some physical quantities of the
LTB-$\Lambda$CDM model from cosmological measurements one can just
use the obtained values of $\Omega_{m}$ and $H_{0}$ from the fit of
the $\Lambda$CDM model (i.e. it is not necessary to impose a special
constraint on the LTB-$\Lambda$CDM model). For example, in
\cite{Kolb} the authors just take $\Omega_m$=0.3 to mimicking a flat
$\Lambda$CDM with mass ratio $\rho_{m0}/\rho_{c0}=0.3$.

\section{Cosmic age at our position}\label{sec:CosmicAge}

In the first subsection, we derive the expression of the cosmic age
in the LTB-$\Lambda$CDM model. Then in the second subsection, we
estimate the age from cosmological observations, and compare it with
that in the homogeneous $\Lambda$CDM model. Finally, in the third
subsection, we discuss the validity of the LTB-$\Lambda$CDM model by
considering the astronomical measurements of the age of some old
objects in our universe.

\subsection{Cosmic Age in the LTB-$\Lambda$CDM}

First of all, we investigate the properties of
Eqs.(\ref{E1}),(\ref{E2}),(\ref{BasicEq.1}-\ref{BasicEq.5}) at $r=0$
(corresponds to our position in the universe). Since the function
$R(r,t)$ represents the diameter distance, one expects
$R(r,t)\rightarrow0$ when $r$ approaches zero. So $R(r,t)$ can be
expanded near $r=0$ as,
\begin{equation}
R(r,t)=R_1(t)r+\ldots
\end{equation}
Then from Eq. (\ref{BasicEq.4}) we obtain
\begin{equation} \label{R1}
\alpha(r)=\alpha_3r^3+....,\ \
\beta(r)=\beta_2r^2+...,\ \
\dot{R}_1(t)=\sqrt{\beta_2+\frac{\alpha_3}{R_1(t)}}.
\end{equation}
From Eqs. (\ref{LCDMValue})(\ref{BasicEq.2})(\ref{BasicEq.4}), we
expand $H_{\Lambda \textrm{CDM}}(z),\ \hat r(z),\ \hat R(z)$ to the
leading order and substitute them into Eq. (\ref{BasicEq.4}) to
obtain $\alpha_3$
\begin{equation}\label{alpha3}
H_{\Lambda \textrm{CDM}}(z)=H_0+...,\ \ r=z+...,\ \ \hat{R}(z)=\frac{z}{H_0}+...,\
\ \alpha_3=\Omega_m H^{-1}_0.
\end{equation}
The calculation of $\beta_2$ is also straightforward. From
Eqs.(\ref{BasicEq.4})(\ref{BasicEq.5}) it follows that
\begin{equation} \label{beta2}
\beta(r)=\big(  \frac{1}{2} \big[ \frac{1}{H_0d\hat{R}/dr}\big(
1-\frac{\alpha}{\hat R}\big) +H_0\frac {d\hat R}{dr}{}  \big]
\big)^2-1.
\end{equation}
Substituting
$\hat{R}^{\prime}(z)=H^{-1}_0+\hat{R}^{\prime}_1z+\hat{R}^{\prime}_2z^2+...$
into Eq. (\ref{beta2}), we obtain
\begin{equation} \label{beta2two}
\beta(z)=(H^2_0\hat{R}^{\prime 2}_1 -
\Omega_m)z^2+...=(H^2_0\hat{R}^{\prime 2}_1 - \Omega_m)r^2+...
\end{equation}
Notice that $\hat R^{\prime}_2$ does not appear in the expression,
so we just have to expand $\hat R^{\prime}$ to the first order.
Using Eqs.(\ref{BasicEq.1})(\ref{BasicEq.3}), it follows that,
\begin{equation}
\frac{d\hat{R}}{dr} = \frac{d\hat{R}}{dz} \frac{dz}{dr} = -
\frac{1}{1+z} \frac{H_{\Lambda \textrm{CDM}}(z)}{H_0} \int^z_0
\frac{dz_1}{H_{\Lambda \textrm{CDM}}(z_1)}+\frac{1}{H_0} = H^{-1}_0
- H^{-1}_0 z+...,\ \ \Rightarrow\ \ \ \ \ \   \hat
R^{\prime}_1=H^{-1}_0.
\end{equation}
Combining with Eq. (\ref{beta2two}) we obtain
\begin{equation} \label{BasicValue2}
\beta_2=1-\Omega_{m}.
\end{equation}
Next we calculate the age of the universe ($t_0-t_{BB}$) at $r=0$.
From Eq. (\ref{R1}) it follows that,
\begin{equation} \label{CosmicAge1}
t_0-t_{BB}=\int^{t_0}_{t_{BB}}dt=\int^{R_1(t_0)}_{R_1(t_{BB})}
\frac{dR_1(t)}{\dot{R}_1(t)},
\end{equation}
where the upper and lower bounds of the integral can be determined
by Eq. (\ref{E2}),
\begin{equation}
R_1(t)=\big( \frac{3\alpha_3}{\kappa \rho(0,t)}  \big)^{\frac{1}{3}}.
\end{equation}
Finally, combining the above with
Eqs.(\ref{LCDMValue})(\ref{alpha3}), we obtain the age of the
universe at $r=0$ in the LTB-$\Lambda$CDM model,
\begin{equation}\label{CosmicAge2}
t_0-t_{BB}=\int^{H^{-1}_0}_{0} \frac{dR_1}{\sqrt{
\Omega_{\Lambda}+\Omega_{m}H^{-1}_0/R_1 }}.
\end{equation}

\subsection{Estimate $t_{LTB}$ from Cosmological Observations}

Next we estimate the cosmic age in the LTB-$\Lambda$CDM model and
compare the result with that of the $\Lambda$CDM model. These two
models are both determined by two parameters, the present matter
ratio $\Omega_m$, and the Hubble constant $H_0$. For convenience,
let us denote the cosmic age in the LTB-$\Lambda$CDM model and
$\Lambda$CDM model at $r=0$ as $t_{LTB}(\Omega_m,H_0)$ and
$t_{\Lambda \textrm{CDM}}(\Omega_m,H_0)$. We have [we perform a
parameter transformation $r=H^{-1}_0/(1+z)$ in $t_{\Lambda
\textrm{CDM}}$]
\begin{eqnarray}
t_{LTB}(\Omega_m,H_0)=\int^{H^{-1}_0}_{0} \frac{dR_1}{\sqrt{
\Omega_{m}/(H_0R_1)+1-\Omega_{m} }}, \ \ \ \ \ \ \ \ \ \ \ \ \ \ \ \ \  \\
t_{\Lambda \textrm{CDM}}(\Omega_m,H_0)=\int^{+\infty}_0
\frac{dz}{H_{\Lambda}(z)(1+z)}=\int^{H^{-1}_0}_0
\frac{dr}{\sqrt{\Omega_m/(H_0r)+(1-\Omega_m)(H_0r)^2}}.
\end{eqnarray}
The only difference between $t_{\Lambda \textrm{CDM}}$ and $t_{LTB}$
lies in the second term in the square root. Since $H_0r<1$ (the
upper bound of the integral is $H^{-1}_0$), it is clearly that
$t_{LTB}$ is smaller than $t_{\Lambda \textrm{CDM}}$ given the same
set of parameters.

Notice that $t_{LTB}(\Omega_m,H_0)$ is proportional to $1/H_0$. To
prove this one can perform a coordinate transformation
$\tilde{R}_1=H_0R_1$, which yields that,
\begin{equation}
t_{LTB}(\Omega_m,H_0)=\frac{1}{H_0}\int^1_0
\frac{d\tilde{R}_1}{\sqrt{ \Omega_{m}/\tilde{R}_1+1-\Omega_{m} }}.
\end{equation}
So we have,
\begin{eqnarray}
\frac{\partial}{\partial H_0} t_{LTB}(\Omega_m,H_0) &=&
-\frac{1}{H_0}t_{LTB}(\Omega_m,H_0)\ <\ 0,\ \ \ \  \\
\frac{\partial}{\partial \Omega_m} t_{LTB}(\Omega_m,H_0) &=&
-\frac{1}{2}\int^{H^{-1}_0}_0 \big(\
\frac{\Omega_m}{H_0r}+1-\Omega_m
\big)^{-3/2}\big(\frac{1}{H_0r}-1\big)\ dr <\ 0,
\end{eqnarray}
Clearly, $t_{LTB}(\Omega_m,H_0)$ has larger values with smaller
values of $\Omega_m$ and $H_0$ parameters. One can easily verify
that the conclusion is the same for $t_{\Lambda
\textrm{CDM}}(\Omega_m,H_0)$.

Now we are ready to calculate the specific values of $t_{LTB}$ with
given values of $H_0$ and $\Omega_m$. Here we refer to the result of
the seven-year Wilkinson microwave anisotropy probe (WMAP)
observations \cite{WMAP7}. From the constraint from
``WMAP7+BAO+$H_0$" \cite{WMAP7}\cite{Riess:H}\cite{BAO2} the WMAP
collaboration provides the best-fit values of $\Omega_{\Lambda}$ and
$H_0$ together with their 1$\sigma$ uncertainties,
\begin{equation}
\Omega_{\Lambda}=0.728^{+0.015}_{-0.016},\ \ H_0=70.4^{+1.3}_{-1.4}
\ \textrm{km/s/Mpc}.
\end{equation}
From their result we can put a constraint on $t_{LTB}$ and
$t_{\Lambda \textrm{CDM}}$ at 1$\sigma$ confidence level (CL). The
result is,
\begin{equation}
t_{LTB}=11.4 \pm 0.3\ \textrm{Gyr},\ \ \ t_{\Lambda
\textrm{CDM}}=13.8 \pm 0.5\ \textrm{Gyr}.
\end{equation}
It is found that $t_{LTB}$ is about 2Gyr younger than $t_{\Lambda
\textrm{CDM}}$. At the 1$\sigma$ CL we obtain the upper limit of the
age of the LTB-$\Lambda$CDM model $t_{LTB}<11.7$Gyr, with the set of
the smallest values of parameters $\Omega_m=0.257$ and
$H_0=69.0$km/s/Mpc (In fact this result is overestimated since we
ignore the degeneracy between $\Omega_m$ and $H_0$. The upper limit
value of $t_{LTB}$ should be smaller, or at least as small as
11.7Gyr).

It should be stressed that to be strict the previous estimation of
$t_{LTB} $ is not appropriate, since we have assumed that the WMAP
and baryon acoustic oscillations data could be used to constrain the
LTB models. In fact, the issues of CMB and structure formation in
the LTB scenario are rather complicated and have not been clearly
investigated. To avoid this problem one can put constraints to
$\Omega_m$ and $H_0$ from SNIa observations. The recent observations
of SDSS-II (Sloan Digital Sky Survey II)\cite{SDSS2} and Hubble
Space Telescope\cite{Riess:H} show that $\Omega_m>0.224,\
H_0>70.6$km/s/Mpc in 1$\sigma$ CL. From their result we find a upper
limit $t_{LTB}<11.6$Gyr, which is the similar with our previous
result $t_{LTB}<11.7$Gyr.

\subsection{Discussions of the Validity of the LTB-$\Lambda$CDM Model}

The low limit to the cosmic age can be directly obtained from
estimating the age of some old objects in our universe
\cite{Richer}\cite{Hansen}\cite{Krauss}. As an example, based on
white dwarf cooling the authors of \cite{Hansen} get a result of
$12.7 \pm 0.7$Gyr. Compared with the result of the previous
subsection $t_{LTB}<11.7$Gyr it seems that the LTB-$\Lambda$CDM
model is inconsistent with their measurements. However, one should
not conclude hastily. The reason is that the result of \cite{Hansen}
is subject to larger uncertainty. The uncertainty due to
calculations of the white dwarf cooling is difficult to estimate,
and in their result corresponding errors are not included. In fact,
the authors of \cite{Hansen} argued that systematic uncertainties
are likely to be at least as large as, if not larger than, the
quoted statistical errors. So if indeed the uncertainty due to
calculations of the white dwarf cooling is as large as the
observational error, then the LTB-$\Lambda$CDM model is within
1$\sigma$ agreement with observations.

The age based on evolution of compact binaries is somewhat lower. In
\cite{Kaluzny}\cite{Chaboyer} the authors give a result of $11.8 \pm
0.6$Gyr and $11.10 \pm 0.67$Gyr, respectively. Moreover, the oldest
known star in the Milky Way, HE 1523-0901, is reported to have an
age of 13.2$\pm$2.7Gyr \cite{HE1523}, for a lower limit of 10.5Gyr.
Obviously, these measurements are all in consistent with $t_{LTB}$
and $t_{\Lambda \textrm{CDM}}$ obtained with data from \cite{WMAP7}
at 1$\sigma$ CL. Therefore, although the obtained $t_{LTB}$ is about
2Gyr younger than $t_{\Lambda \textrm{CDM}}$, it is still in
1$\sigma$ agreement with current astronomical observations, and we
are not able to argue against the reconstructed LTB-$\Lambda$CDM
model.

We show the situation in Fig. 1. The green and red regions represent
parameters with cosmic age older than 11.2Gyr in $\Lambda$CDM and
LTB-$\Lambda$CDM model, respectively. The black shadow region is a
1$\sigma$ CL constraint from the seven-year WMAP observations
\cite{WMAP7} (we ignore degeneracy). The blue region is a 2$\sigma$
CL constraint from a joint analysis performed in one of our previous
works \cite{Constraints}, in which we used the Constitution
supernovae sample \cite{Constitution}, the baryon acoustic
oscillations \cite{BAO} and the five-year WMAP observations
\cite{WMAP5}. Since current limit to the cosmic age from
astronomical measurements generally gives a result $t<11.2$Gyr, we
plot the regions of $t_{LTB}<11.2$Gyr (red shadow) and
$t_{\Lambda\textrm{CDM}}<11.2$Gyr (green shadow) in this figure. It
is obvious that the $\Lambda$CDM perfectly passes the cosmic age
text, while the overlap of the blue region, the black shadow region
and the red shadow region implies that the LTB-$\Lambda$CDM model is
also consistent with current observations.

Finally, at the end of this section, we stress that in this paper we
only consider a particular LTB model - namely, the one with
$\Lambda$CDM features. Other inhomogeneous models (such as the void
models) may have larger age at the origin, and in these cases one
would have to seek for other methods to test and identify them.

\begin{figure}\label{Fig1}
\includegraphics[width=12.3cm]{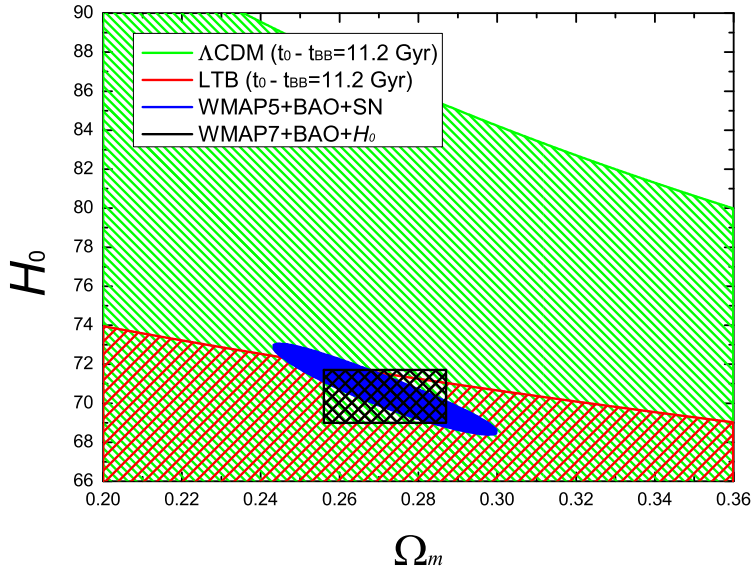}
\caption{Parameter space in $\Omega_m$ and $H_0$ plane. The green
and red regions stand for $t_{\Lambda \textrm{CDM}}>$11.2Gyr and
$t_{LTB}>$ 11.2Gyr respectively. The black shadow region represents
the 1$\sigma$ constraint to $\Omega_m$ and $H_0$ from the seven-year
WMAP observations (degeneracy is ignored). The blue region is a
2$\sigma$ constraint from a joint analysis from the Constitution
supernovae sample, baryon acoustic oscillations  and the five-year
WMAP observations. }
\end{figure}

\section{Summary}\label{sec:Summary}

In this paper we calculate the cosmic age at $r=0$ in the
LTB-$\Lambda$CDM model, which reproduces the luminosity-distance and
light-cone matter density of the homogeneous $\Lambda$CDM model.
Using the constraints of $\Omega_m$ and $H_0$ from the seven-year
WMAP observations combined with other cosmological observations, we
get the upper limit $t_{LTB}<11.7$Gyr at 1$\sigma$ CL. This result
is about 2Gyr younger than the cosmic age in $\Lambda$CDM scenario.
Since current astronomical measurements generally put a 1$\sigma$ CL
lower limit on the age of the universe of about 11.2Gyr, the
LTB-$\Lambda$CDM model is still in 1$\sigma$ agreement with all
these observations. However, due to the relatively younger age the
LTB-$\Lambda$CDM model might be disfavored by future observations.

Besides, even if the LTB-$\Lambda$CDM model successfully passes all
the tests of future observations, there might be some other problems
in this scenario.(The discussions of these complicated problems are
beyond the scope of this paper.) The main reason is that it is
difficult to fit this model into the larger framework of fundamental
physics such as particle physics, general relativity, astrophysics,
and cosmology. For example, in \cite{Kolb} the authors mentioned the
theory of structure formation and the integrated Sachs-Wolfe effect:
study of these issues is very difficult in the LTB models.

The topic of distinguishing the homogeneous $\Lambda$CDM model and
the reconstructed inhomogeneous LTB-$\Lambda$CDM model is
scientifically interesting and important, since it involves the
question of the mysterious feature of dark energy and whether the
nearby region of the universe is homogeneous. This topic should be
carefully investigated. In this paper we propose the possibility of
distinguishing the LTB-$\Lambda$CDM scenario with the standard
$\Lambda$CDM model by performing the cosmic age test. The procedure
is convenient and straightforward. Although the result shows that
the LTB-$\Lambda$CDM is still in 1$\sigma$ agreement with current
astronomical observations, since with the same set of parameters
this model always has a younger age than the standard $\Lambda$CDM
model it is possible to distinguish them from future observations.
In all, the issue of using the cosmic age test to distinguish the
reconstructed inhomogeneous LTB models from the homogeneous
$\Lambda$CDM model is worth further investigation, and should be
taken into consideration in future works, e.g., in the cases when
people try to construct a new model in the LTB scenario to explain
the apparent cosmic acceleration.

\begin{acknowledgments}
The authors would like to thank the anonymous referees for carefully
examining our paper and providing us a number of important comments.
This work was supported by the Natural Science Foundation of China
under Grants No. 10535060/A050207, No. 10975172 and No. 10821504,
and Ministry of Science and Technology 973 program under Grant No.
2007CB815401. SW also thanks the support from a graduate fund of the
University of Science and Technology of China.
\end{acknowledgments}


\end{document}